\documentstyle[prc,aps,twocolumn,epsfig]{revtex}

\newcommand{\vsig}{\mbox{\boldmath $\sigma$ \unboldmath}}
\newcommand{\veps}{\mbox{\boldmath $\epsilon$ \unboldmath}}
\newcommand{\valf}{\mbox{\boldmath $\alpha$ \unboldmath}}

\begin{document}

\title{ Quark model predictions for $K^*$ photoproduction
on the proton }

\author{Qiang Zhao$^1$\thanks{Electronic address: 
qiang.zhao@surrey.ac.uk.}, J.S. Al-Khalili$^1$, and C. Bennhold$^2$}
\address{1) Department of Physics, University of Surrey, Guildford, Surrey, 
GU2 7XH, United Kingdom}
\address{2) Department of Physics, Center for Nuclear Studies,\\
 The George Washington University, Washington, D.C., 20052, USA}

\date{\today}

\maketitle  
  
\begin{abstract}
The photoproduction of $K^*$ vector mesons is investigated
in a quark model with an effective Lagrangian.
Including both baryon resonance excitations and {\it t}-channel exchanges, 
observables for the reactions $\gamma p\to K^{*0}\Sigma^+$ and $\gamma p\to
K^{*+}\Sigma^0$ are predicted, using the
SU(3)-flavor-blind assumption of non-perturbative QCD.
\end{abstract}

\vskip 0.5cm

PACS number(s): {\small 12.39.-x, 13.60.-r, 13.60.Le, 14.20.Gk }

\vskip 0.5cm

\section{ Introduction}

The availability of
high-intensity photon and electron facilities at JLAB, ELSA, ERSF and
SPring-8 has revived both experimental and theoretical interest 
in searching for
``missing resonances"~\cite{NRCQM-1,NRCQM-2} 
in meson photo- and electroproduction processes. 
Baryon resonances excited by electromagnetic probes can be
investigated through their various meson-nucleon and meson-hyperon decay
channels. The exclusive study of meson photo- or electroproduction
provides insights into the internal structures of the intermediate
states.  For $K^*$ photoproduction JLAB and ELSA have recently
taken the first ever exclusive measurements in the resonance region.

The photoproduction of $K^*$ vector mesons intersects on the one hand
with other strangeness production reactions, such as $\gamma p\to K \Lambda$ 
and
$\gamma p\to K \Sigma$, and on the other hand with the field of non-strange
vector meson production. Much work has been done in recent years on
the former 
reactions~\cite{bennhold-nucl-th-9901066,feuster1999,David:1996pi}.
At quark level, both
$(\gamma, K)$ and $(\gamma, K^*)$ contain clear information about 
the production
of an $s \overline{s}$ pair from the vacuum.  At the hadronic level
these reactions are related to each other 
since one reaction contains the meson produced in the other one 
as the {\it t}-channel exchanged particle, thus constraining the range of
available couplings.  With regard to the field of non-strange vector 
meson photoproduction, 
only recently have there been theoretical efforts devoted 
to the processes $\gamma N \to \rho N$ and $\gamma N \to \omega N$
in the resonance region~\cite{plb98,prc98,zhao-omega,oh2000}.  Sharing
the same observables, the $(\gamma,K^*)$ process can benefit
from work on the structure of and the relationship between
helicity amplitudes, vector meson multipoles and polarization 
observables~\cite{schilling,tabakin}.
In contrast to the $\gamma p \to \rho^0 p $, $\gamma p \to \omega p$ and
$\gamma p \to \phi p$ reactions, 
the {\it t}-channel Pomeron exchange is not possible
for the $\gamma p \to K^* Y$ process, where $Y$ denotes the hyperon. 
In this respect it shares similar features
with charged $\rho$ photoproduction, thus simplifying the study of
intermediate resonance excitation.

The reason we choose a quark model approach is to avoid the 
uncertainties arising from a lack of knowledge about the $K^*\Sigma N^*$
couplings. A recent study of $N^*\to K^* Y$ by Capstick and
Roberts~\cite{capstick98}, using a quark-pair-creation model,
suggests that 
most $K^*Y$ decay branching ratios are small due to the high threshold for
these channels. 
Only a few low-lying negative-parity states were predicted to be
strongly coupled to the $K^*\Sigma$ channels, including $N[\frac 12
^-]_5(2070)$ (established in pion production), $\Delta[\frac 32
^-]_3(2145)$, and $\Delta[\frac 12 ^-]_3(2140)$.
Note that only those resonances above the decay threshold
can be predicted by Ref.~\cite{capstick98}, 
the vector meson production could be the place where those
below-threshold-resonance
couplings to $K^*Y$ can be investigated.
In our model, 
apart from the commonly-used quark
model parameters, there are only two free parameters relating to the
$K^*\Sigma N^{(*)}$ couplings that appear in the quark model symmetry limit.
These are the vector and tensor couplings for the quark-$K^*$
interaction.  Meanwhile,
the SU(3)-flavor-blind assumption is made
in this model and suggests that these parameters should have values close to
those used in the $\omega$ and $\rho$ meson photoproductions.  
In this study, we shall adopt the most recently extracted information 
from the $\omega$ meson photoproduction as an input.

In this work, we present quark model predictions for the $K^{*0}$ and
$K^{*+}$ photoproductions: $\gamma p\to K^{*0}\Sigma^+$ and
$\gamma p\to K^{*+}\Sigma^0$.  It is the first theoretical attempt to
study nucleon resonance excitations in these two channels.  
Low-lying resonances within the $n\le 2$ harmonic oscillator shells
will be included explicitly in the formalism, while higher-mass
resonances with $n>2$, for which there is little information, are
treated as being degenerate by summing over all states for the same $n$,
which are taken to have the same mass and width~\cite{zpli97}. 
As listed in the
Particle Data Group~\cite{PDG00}, baryon resonances above 2 GeV
generally have widths around 300-400 MeV, and large mass-overlaps make
this approximation a reasonable one. Taking such a scheme, 
we will concentrate in this paper
on the following points:
(i) the magnitude of the differential and total cross sections 
predicted by such a model,
(ii) the differences between neutral and charged $K^*$ photoproduction and
(iii) the effects of a {\it t}-channel 
kaon exchange in analogy with our study of the non-strange 
$\omega$ and $\rho$ meson photoproductions~\cite{plb98,prc98,zhao-omega}.
These results will be tested by forthcoming data 
from JLAB and ELSA in the near future.

This paper is arranged as follows. In Section II, the $K^*$ photoproduction
formalism with an effective Lagrangian is introduced. Section III
presents the kaon exchange contributions for both reactions. 
Numerical results and discussions are given in Section IV.

\section{$K^*$ photoproduction at tree level}

The $K^*$ meson photoproduction involves the creation of $s$ and $\overline{s}$
quarks in the SU(3) quark model.  
With the effective Lagrangian of Refs.~\cite{plb98,prc98}
adopted for the $K^*$-$qq$ vertex, the $s$ quark
will couple to the meson in the same way as the $u$ and $d$ quarks,
 apart from its
different mass ($m_s=500$ MeV, $m_u=m_d=330$ MeV).
That is to say, the SU(3)-flavor-blind assumption of non-perturbative QCD
is adopted in our model.

Following the convention of Ref.~\cite{prc98}, 
the transition amplitude can be expressed 
as the sum over the {\it t}-, {\it s}- 
and {\it u}-channels, 
\begin{equation}
\label{amplitude}
M_{fi}=M^t_{fi} + M^s_{fi} + M^u_{fi} \ , 
\end{equation}
where the resonance excitations enter the
{\it s}- and {\it u}-channels explicitly as follows, 
\begin{eqnarray}  
\label{s+u}
&&M^s_{fi}+M^u_{fi}\nonumber\\
&=  &\sum_{j} \langle N_f|H_m|N_j\rangle\langle   
N_j|\frac{1}{E_i+\omega_\gamma-E_j}H_{em}|N_i\rangle\nonumber\\  
&&+\sum_{j} \langle N_f|H_{em}\frac{1}{E_i-\omega_m-E_j}
|N_j\rangle\langle N_j|H_m|N_i\rangle\nonumber\\
&= &i\langle N_f|[g_e, H_m]| N_i\rangle \nonumber\\
&& +i\omega_\gamma \sum_{j} \langle N_f|H_m|N_j\rangle\langle   
N_j|\frac{1}{E_i+\omega_\gamma-E_j}h_e|N_i\rangle\nonumber\\  
&& +i\omega_\gamma \sum_{j} \langle N_f|h_e\frac{1}{E_i-\omega_m-E_j}
|N_j\rangle\langle N_j|H_m|N_i\rangle , 
\end{eqnarray}  
with the quark-meson coupling,
\begin{equation}\label{lagrangian}
H_m=-\overline{\psi}_l(a\gamma_\mu +\frac{ib}{2m_q}
\sigma_{\mu\nu}q^\nu) V^\mu \psi_l \,
\end{equation}
and the quark-photon electromagnetic interaction,
\begin{equation} 
H_{em}=-\overline{\psi}_l\gamma_\mu e_l A^\mu\psi_l \,
\end{equation}
where $\psi_l$ ($\overline{\psi}_l$) is the quark (anti-quark)
field, and $V^\mu$ is the vector meson field; 
$e_l$ is the charge operator for the $l$th quark; 
$\omega_\gamma$ and $\omega_m$ denote the 
energy of the incoming photon and outgoing meson
in the c.m. system, respectively, while $E_i$ and $E_j$
are the energy of the initial state proton and intermediate 
baryons (or hyperons). The summation should run 
over all the intermediate
states $|N_j\rangle$, which can be explicitly 
described by the symmetric quark model wavefunctions.

In the second equivalence of Eq.~(\ref{s+u}),
a standard transformation 
has been used (see Eqs.~(14)-(19) in Ref.~\cite{zpli97}
for explicit deduction), where
the electromagnetic interaction is transformed
into, $g_e\equiv \sum_{l} e_l {\bf r}_l\cdot\veps e^{i{\bf k}\cdot{\bf r}_l}$,
and $h_e\equiv \sum_{l} e_l {\bf r}_l\cdot\veps 
(1-\valf\cdot{\bf k}/\omega_\gamma ) e^{i{\bf k}\cdot{\bf r}_l}$.
Thus, we re-define
the second and third term as
the {\it s}- and {\it u}-channel transition, respectively,
while the first term is identified as a ``seagull" term. 
In Ref.~\cite{prc98}, it was shown that the seagull term 
accompanied by the {\it t}-channel transition amplitude
was essential for recovering gauge invariance for the theory.

An interesting feature arising from the effective Lagrangian
is that 
the {\it t}-channel vector meson exchange 
$M^t_{fi}$ and the seagull term 
are proportional to the 
charge of the outgoing $K^*$ meson. Thus, 
they will vanish in the neutral vector meson production 
but play a role as background
in the charged meson production. Meanwhile, 
the diffractive process is absent in the $K^*$ photoproduction  
due to the strangeness production. 
These two features together make the neutral $K^{*0}$ production
an opportunity to study the {\it s}- and {\it u}-channel 
contributions in the absence of those nonresonant processes.

The two parameters, $a$ and $b$ in Eq.~(\ref{lagrangian}), 
represent the vector and 
tensor couplings at the quark-meson interaction vertex, and 
are the basic parameters in this model.
At present, no experimental data for $K^*$ photoproduction
can be used to constrain them.  
As mentioned above, the SU(3)-flavor-blind approach
suggests that apart from the different quark mass, the
parameters $a$ and $b$ should have the same magnitudes as those
derived in the photoproduction of the non-strange
$\omega$ and $\rho$ mesons. Therefore, we shall 
adopt the parameters used in the
$\omega$ meson photoproduction~\cite{zhao-omega} as a first test.

\section{ Kaon exchange terms }

Our study of $\omega$ and $\rho^0$  meson photoproduction
showed that the {\it s}- and {\it u}-channel contributions
from the effective Lagrangian are not sufficient to describe the corresponding 
vector meson photoproduction reactions. 
Apart from diffractive Pomeron exchange
in neutral vector meson production~\cite{prc98,zhao-omega}, 
scalar or pseudoscalar
meson exchange terms are needed to reproduce the small angle forward-peaking
behavior. For example, in $\gamma p \to \omega p$, the $\pi^0$ exchange 
dominates at small angles and accounts for large fractions of the cross
section from threshold up to around 2.2 GeV. With the pion exchange
as the main {\it unnatural}-parity exchange process, it
plays an important role in parity asymmetry observables. 
Similarly, in the reaction $\gamma p\to \rho^0 p$, a {\it t}-channel 
$\sigma$ meson exchange must be introduced as well. 

On the other hand, studies of kaon photoproduction with isobaric
models~\cite{bennhold-nucl-th-9901066,adelseck-saghai-90} showed that
$K^*$ exchange terms were needed to reproduce forward peaking
at higher energies. A chiral quark model study~\cite{zpli-Kaon-95,zpli-Kaon-96} 
arrived at similar conclusions.  Recent data from
SAPHIR~\cite{Goers:1999sw,bennhold-98} for $\gamma p \to K^+ \Sigma^0$ and $\gamma p
\to K^0 \Sigma^+$  clearly showed a forward-peaking in both of these 
channels as well.  This set of data provides interesting information to compare
with the $K^*$ photoproduction, where kaon exchange might play a
similar role, since the $K^*K\gamma$ vertices are the same 
in both processes.

Taking into account the above features in $K$ photoproduction on the one hand
and $\omega$ and $\rho$  photoproduction on the other,
it is reasonable to assume that a {\it t}-channel leading-order
light meson exchange is needed for the $K^*$ meson photoproduction as well.
This kind of information, even if only based on phenomenological
considerations, may help in shedding more light on the hadron duality
hypothesis~\cite{dolen-66,williams}.

The kaon exchange is introduced in with a  
pseudoscalar coupling for the $K\Sigma N$ vertex, which has 
the same form as a pseudovector coupling at tree level,  
\begin{equation}
\label{K-Sigma-N}
{\cal L}_{K\Sigma N}=-i g_{K\Sigma N} \overline{\psi}_\Sigma\gamma_5
 \psi_p \phi_K\ ,
\end{equation}
where $g_{K\Sigma N}$ is the $K\Sigma N$ coupling constant; 
$\psi_p$ and $\overline{\psi}_\Sigma$ are
the initial state proton and final state $\Sigma$ baryon, respectively; 
$\phi_K$ denotes the kaon.
The $K^* K\gamma$ vertex is, 
\begin{equation}
\label{Kaon-decay}
{\cal L}_{K^* K\gamma}=e\frac{g_{K^* K\gamma}}{M_{K^*}}
\varepsilon_{\alpha\beta\gamma\delta}\partial^\alpha A^\beta 
\partial^\gamma V^\delta \phi_K \ ,
\end{equation}
where $A^\beta$ and $V^\delta$ are the photon and vector meson
fields; $g_{K^* K\gamma}$ is the coupling constant;
$M_{K^*}$ denotes the $K^*$ mass.

We determine the $K^* K\gamma$ coupling from $K^*$ radiative decay, 
$K^* \to K + \gamma$. Adopting the partial decay width, 
$\Gamma_{(K^{*0} \to K^0 + \gamma)}=116$ keV and 
$\Gamma_{(K^{*+} \to K^+ + \gamma)}=50$ keV ~\cite{PDG00}, 
we obtain $g_{K^{*0} K^0\gamma}=1.134$ and $g_{K^{*+} K^+\gamma}=0.744$, 
respectively for the neutral and charged decay channels. 
A quark model constraint is adopted for 
the relative sign between these two couplings, 
i.e. a sign difference exists.
For the $K\Sigma N$ coupling, there is more uncertainty over
the value that should be used. 
The most recent studies~\cite{bennhold-nucl-th-9901066} suggest that 
$g_{K^+\Sigma^0 p}/\sqrt{4\pi}=1.2$ can be regarded as a 
reasonable value.~\footnote{It was also found 
in Ref.~\cite{bennhold-nucl-th-9901066} that
a free-parameter-fitting of 
data gave $g_{K^+\Sigma^0 p}/\sqrt{4\pi}=-0.37$, which 
would produce negligible effects in this calculation. }
Isospin symmetry is taken into account and 
gives the following relation,
\begin{equation}
\frac{g_{K^0\Sigma^+ p}}{g_{K^+\Sigma^0 p}}=
\frac{g_A(\gamma p \to K^0 \Sigma^+) }{ g_A(\gamma p \to K^+ \Sigma^0)} 
 = \sqrt{2} \ ,
\end{equation}
where the $g_A$ is the axial vector coupling constant in pseudoscalar 
meson photoproduction. The ratio of the two values of $g_A$  
is derived in the SU(3) quark model. 

From Eqs.~(\ref{K-Sigma-N}) and ~(\ref{Kaon-decay}), 
the {\it t}-channel kaon exchange amplitudes can be 
written as, 
\begin{eqnarray}
M^t_T&=&\frac{e g_{K\Sigma N} g_{K^*K\gamma}}{2M_{K^*}(t-m^2_K)}
e^{-({\bf q}-{\bf k})^2/6\alpha_k^2}\nonumber\\
&&\times\{\omega_\gamma\veps_\gamma\cdot({\bf q}\times\veps_v)
+\omega_m{\bf k}\cdot(\veps_\gamma\times\veps_v)\}
\vsig\cdot {\bf A} \ ,
\label{t}
\end{eqnarray}
for the transverse transition, and
\begin{eqnarray}
M^t_L&=& -\frac{e g_{K\Sigma N} g_{K^*K\gamma}}{2M_{K^*}(t-m^2_K)}
\frac{ M_{K^*}}{|{\bf q}|}e^{-({\bf q}-{\bf k})^2/6\alpha_k^2}\nonumber\\
&&\times(\veps_\gamma\times{\bf k})\cdot{\bf q} 
\vsig\cdot {\bf A} \ ,
\label{l}
\end{eqnarray}
for the longitudinal transition, 
where ${\bf A}={\bf q}/(E_f+M_\Sigma)-{\bf k}/(E_i+M_N)$ and 
$t=(q-k)^2$.  
The factor $e^{-({\bf q}-{\bf k})^2/6\alpha_k^2}$ comes 
from the spatial integral over the initial state nucleon 
and final state $\Sigma$ baryon, and plays the role 
of a form factor for the {\it t}-channel kaon exchange.
The parameter, $\alpha_k$, in the harmonic oscillator potential is 
treated as a free parameter such that 
it will partly take into account the form factor 
at the $K^* K\gamma$ vertex. 
Here we fix $\alpha_k=290$ MeV, which is 
the same as used in $\omega$
meson photoproduction~\cite{zhao-omega}.

\section{Results and Discussions}

We adopt the following values for the two basic parameters in the quark-meson coupling
vertex, $a=-2.8$ and $b=-5.9$, derived from $\omega$ meson 
photoproduction~\cite{zhao-omega} using preliminary 
polarized beam asymmetry data from GRAAL~\cite{graal-omega}. 

Fig.~\ref{fig:(1)} shows the differential cross sections 
for $\gamma p \to K^{*0} \Sigma^+$ (left column)
and $\gamma p \to K^{*+} \Sigma^0$ (right column)
 for four energies, $E_\gamma=1.88$, 2.10, 2.40 and 2.60 GeV. 
The solid curves denote the results
without the {\it t}-channel kaon exchange. 

For the $K^{*0}$ production, due to the absence of the seagull term and 
$K^{*0}$ exchange,
the solid curves represent contributions
from {\it s}- and {\it u}-channel processes, 
and thus reflect the magnitudes of resonance 
excitations.  
For the $K^{*+}$ production, 
although the near-threshold cross section 
is compatible with the $K^{*0}$ production,
strong forward peaking is found above threshold. 
The change of story is due to the dominant
contribution from the seagull term as well as 
the {\it t}-channel $K^{*+}$ exchange, 
which shifts the peaks even much forward.
The flattened angular distributions in both reactions
are produced by the large mass widths 
for higher mass resonances, which generally belong to $n>2$ harmonic
oscillator shells, and are treated as degenerate for each $n$ 
since little is known about them. 
The off-shell low-lying resonances  
contribute only through their wave function tails, 
far from their mass positions. 
Near threshold, the cross section for the $K^{*0}$ production 
slightly increases at large angles which is found to come from
the {\it u}-channel $\Lambda$ and $\Sigma$ transitions.

Compared to the seagull term and 
nucleon pole terms, the resonance contributions belong to higher orders,
and generally have flattened distributions.
Because of this,
the dominance of the seagull term and {\it t}-channel
$K^{*+}$ exchange will put a strong constraint on the parameters
when experimental data are available.

Calculations including the {\it t}-channel kaon exchanges
are represented by the dashed curves in Fig.~\ref{fig:(1)}.
Comparing these two reactions, it is clear that the near-threshold
region is not sensitive to the possible kaon exchange.  
However, above threshold, 
strong forward-peaking is produced by $K^0$ exchange
for the $K^{*0}$ production, while only a small enhancement
occurs at forward angles for the $K^{*+}$ 
due to the $K^+$ exchange.
An interesting feature of the kaon exchanges
in this two reactions is that 
the $K^+$ exchange is relatively suppressed 
in comparison with the $K^0$ exchange.
It can be seen by the couplings. Namely, 
$g_{K^{*0}K^0\gamma}=-1.525g_{K^{*+}K^+\gamma}$, due to the larger
radiative decay width for $K^{*0}$; and
$g_{K^0\Sigma^+ p}=\sqrt{2}g_{K^+\Sigma^0 p}$ 
given by the SU(3) flavor symmetry.
This feature is very similar to the charged $\rho$ meson
production, where the $\pi^\pm$ exchanges were found to be negligible.

Total cross sections for these two reactions are shown in Fig.~\ref{fig:(2)}.
The dotted and solid curves 
denote the predictions with and without the 
{\it t}-channel $K^+$ exchange for $\gamma p \to K^{*+} \Sigma^0$,
while the dot-dashed and dashed curves denote 
with and without the 
{\it t}-channel $K^0$ exchange for $\gamma p \to K^{*0} \Sigma^+$.
Again we find that cross sections for $K^{*0}$ photoproduction 
are much lower than for $K^{*+}$ production. 
The effects of {\it t}-channel kaon exchange are 
significant for $K^{*0}$ photoproduction, 
but rather small for $K^{*+}$ production. 

\begin{figure}
\caption{ 
Beam polarization asymmetry $\check{\Sigma}_A$ at two energies,
$E_\gamma=1.88$ and 2.10 GeV.
The dashed and 
solid curves denote calculations with and 
without the kaon exchange, respectively.
}
\protect\label{fig:(3)}
\end{figure}
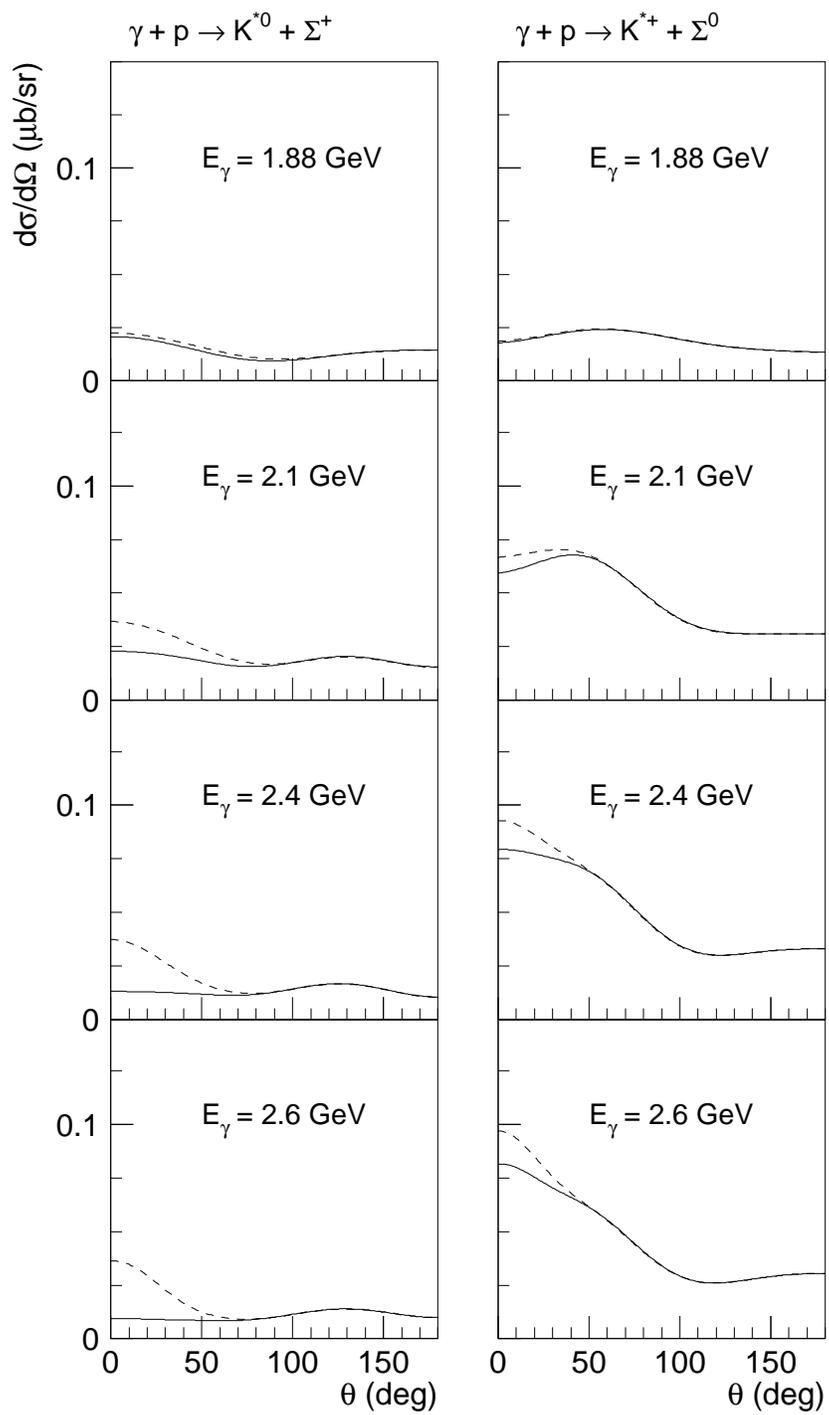
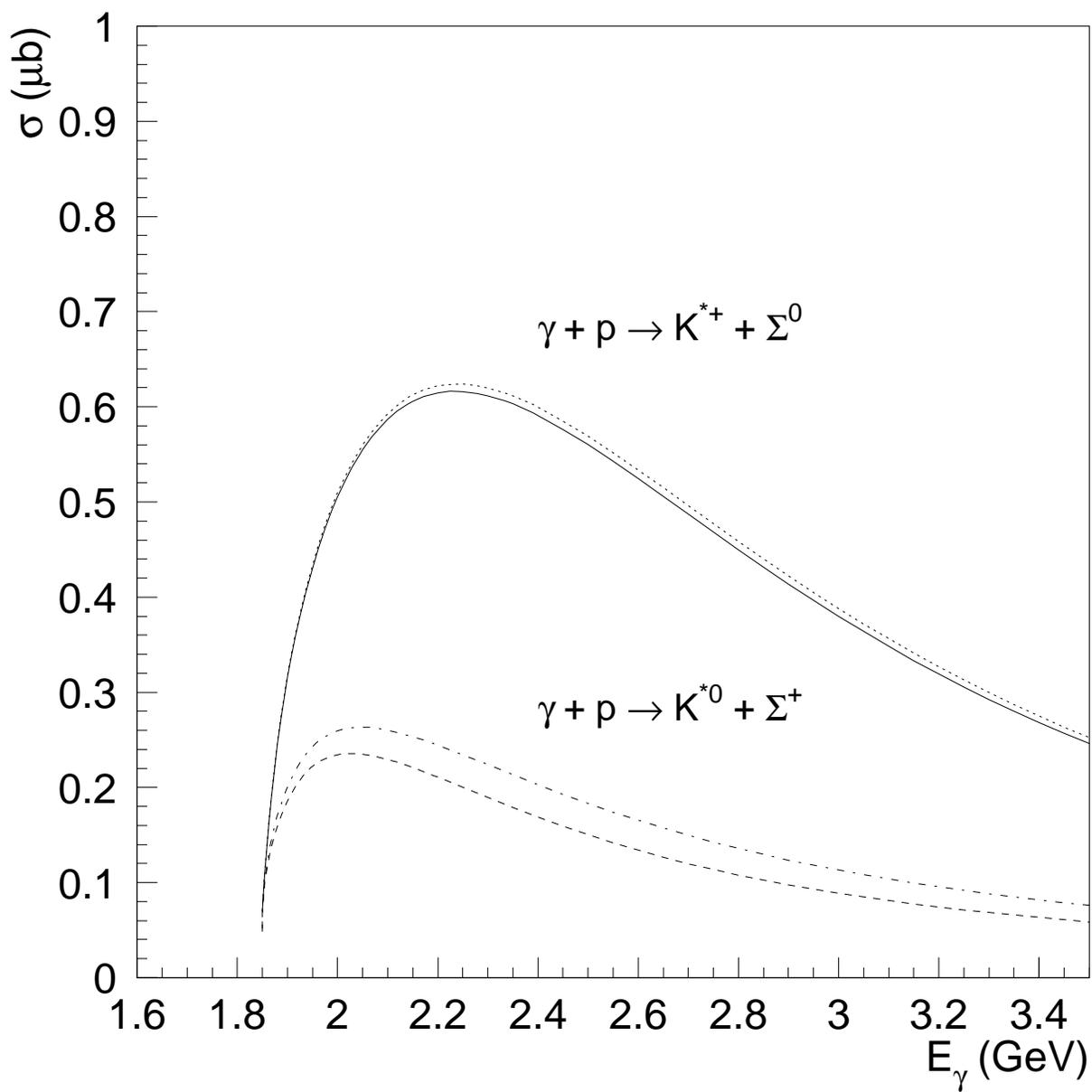
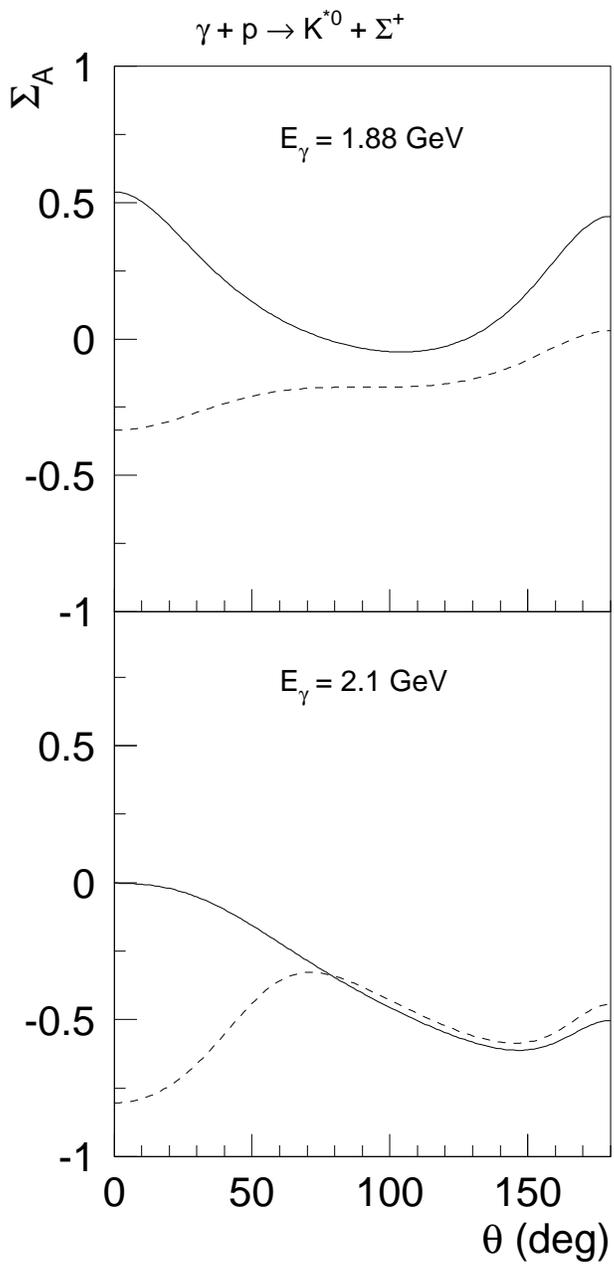
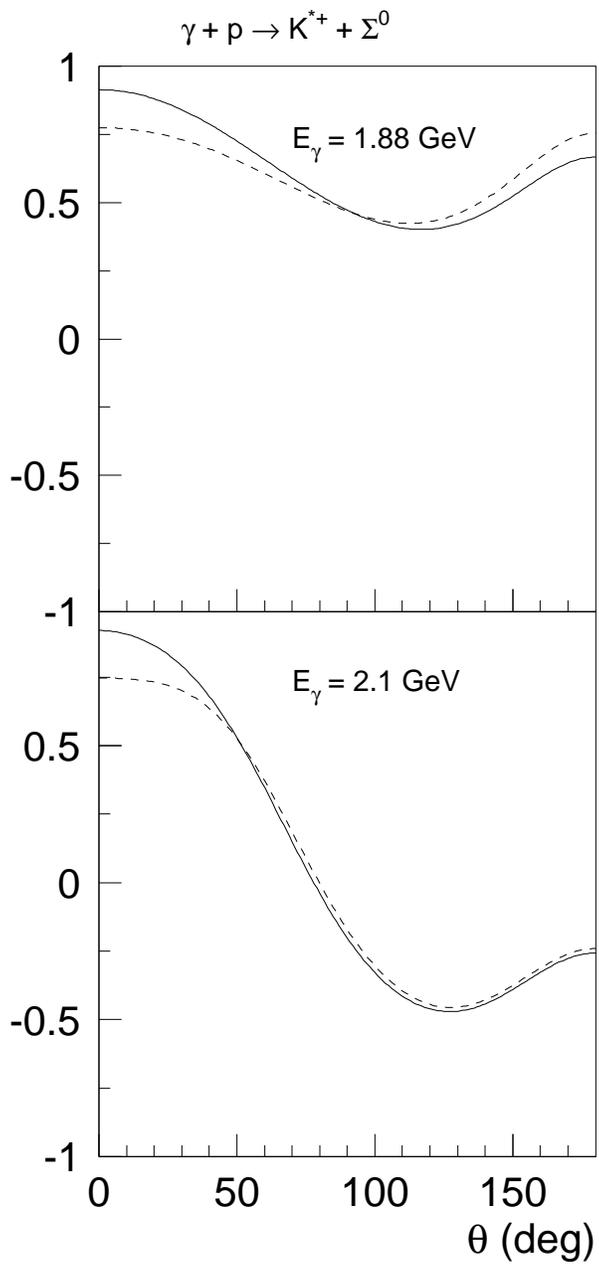

Next we study the influence of {\it t}-channel kaon exchange on the 
beam polarization asymmetries. These {\it unnatural}
parity exchanges and their interferences with the {\it s}- and 
{\it u}-channel transitions are expected to
 be particularly important at forward angles.
Following the convention of Ref.~\cite{schilling},
the beam polarization asymmetry can be expressed, in terms 
of the density matrix elements $\rho^\alpha_{\lambda\lambda^\prime}$ 
of the vector meson decay, as
\begin{equation}
\label{beam-asymm-A}
\check{\Sigma}_A\equiv\frac{\rho^1_{11}+\rho^1_{1-1}}
{\rho^0_{11}+\rho^0_{1-1}} = \frac{\sigma_\parallel -\sigma_\perp}
{\sigma_\parallel +\sigma_\perp} \ ,
\end{equation}
where $\sigma_\parallel$ represents the cross section 
of the vector meson decay with the decay particles in the photon polarization plane, 
while $\sigma_\perp$ represents the cross section with the decay plane 
perpendicular to it.
In Fig.~\ref{fig:(3)}, $\check{\Sigma}_A$ is calculated 
at two energies, $E_\gamma=1.88$ and 2.10 GeV.  
Comparing the solid and dashed curves, 
one can see that this observable is very sensitive to the 
kaon exchanges, whose exclusive contribution 
would result in $\check{\Sigma}_A=-1$. 
For $K^{*+}$ production, small contributions
from the $K^+$ exchange produce significant effects 
at forward angles that shift the asymmetries to smaller values. 
For $K^{*0}$ production, the effect is relatively strong and 
changes the sign of the asymmetry at $E_\gamma=1.88$ GeV. 
Thus, with increasing energy, the 
large angle region will be dominated by the transitions 
from the effective Lagrangian, while the forward angle
asymmetries are determined by the $K^0$ exchange.

In summary, this paper presents the first $K^*$ meson photoproduction
calculations for both isospin
channels, $\gamma p \to K^{*0} \Sigma^+$ and $\gamma p \to K^{*+}
\Sigma^0$, using a quark model with effective Lagrangian.
Overall, the predicted cross sections are much smaller by at least
an order of magnitude compared to either 
$K$ or $\rho$ and $\omega$ photoproduction.
Adopting  quark-meson vector and tensor couplings, $a=-2.8$
and $b=-5.9$, we find that $K^{*+}$ production is significantly larger than
$K^{*0}$ production due to the presence of the seagull term
and the {\it t}-channel $K^{*+}$ exchange contribution.
The effects of pseudoscalar kaon exchange are studied in both reactions. 
We find that using a standard value for the
 $K\Sigma N$ coupling will produce clear forward-peaking
behavior in $\gamma p \to K^{*0} \Sigma^+$, while only 
small enhancements are found in $\gamma p \to K^{*+} \Sigma^0$.
This sensitivity of $K^{*0}$ production to {\it t}-channel $K^0$ exchange
might provide an additional constraint on the $K\Sigma
N$ coupling, a valuable ``by-product" from the measurement of
$K^*$ photoproduction.  
We note that at present the sign between
the kaon exchange terms and the {\it s}- and {\it u}-channel 
$K^*$ production terms is unknown, due to our 
lack of knowledge of the $K^*\Sigma N$ 
couplings. Spin observables, rather than the differential cross section,
would be more sensitive to this interference.

Also we note that this sensitivity is almost independent on the 
degenerate approximation for the $n>2$ states since 
their contributions are generally small. Although 
the quark-$K^*$-meson couplings are 
compatible with the non-strange-quark-meson couplings, 
suppressions of higher partial resonances from the quark model 
form factors and large mass overlapping effects 
result in small cross sections for the $n>2$ terms.
Meanwhile, since
the forward-peaking {\it t}-channel dominates at small angles,
while the {\it s}- and {\it u}-channels (excluding the seagull term)
generally have flattened behavior and dominate at large angles, 
the degenerate approximation turns out to be reasonable.
At this stage.
the calculations presented here should be regarded as a first step
that provides a description of collective resonance excitations, rather
than showing effects arising from individual resonances. 
With regard to our original motivation of searching for ``missing
resonances", future work
needs to include the high-lying states individually in order to 
assess their importance in these reactions. Also, a study of
the SU(6)$\otimes$O(3) symmetry violation would be necessary.

\acknowledgments
The authors thank Z.-P. Li for valuable comments on this work.

%
%
\begin{figure}
\caption{ 
Differential cross sections for $\gamma p \to K^{*0} \Sigma^+$ 
(left column) and $\gamma p \to K^{*+} \Sigma^0$ (right column) 
at four different energies. The dashed and solid curves are 
predictions 
with and without {\it t}-channel kaon exchange, respectively. }
\protect\label{fig:(1)}
\end{figure}
\begin{figure}
\caption{ 
Total cross sections for $\gamma p \to K^{*0} \Sigma^+$ 
with (dot-dashed) and without (dashed) the {\it t}-channel 
$K^0$ exchange, 
and for $\gamma p \to K^{*+} \Sigma^0$ 
 with (dotted) and without (solid) the {\it t}-channel 
$K^+$ exchange. 
}
\protect\label{fig:(2)}
\end{figure}
\end{document}